\documentclass[12pt]{article}
\usepackage[margin=2cm]{geometry}
\usepackage{setspace}
\usepackage[numbers,sort&compress]{natbib}

\usepackage{graphicx}% Include figure files
\usepackage{xcolor}
\usepackage{dcolumn}% Align table columns on decimal point
\usepackage{bm}% bold math
\usepackage{amsmath,amssymb}
\usepackage{hyperref}% add hypertext capabilities
\usepackage{xr}
\makeatletter
\newcommand*{\addFileDependency}[1]{% argument=file name and extension
  \typeout{(#1)}
  \@addtofilelist{#1}
  \IfFileExists{#1}{}{\typeout{No file #1.}}
}
\makeatother
 
\newcommand*{\myexternaldocument}[1]{%
    \externaldocument{#1}%
    \addFileDependency{#1.tex}%
    \addFileDependency{#1.aux}%
}
\myexternaldocument{supp_info_revised_nanoscale}

\usepackage{mathtools}

\newcommand{\XYZ }{ }
\definecolor{myblue}{RGB}{17, 30, 108}
\definecolor{myred}{RGB}{184, 15, 10}

\title{Parametric study of temperature distribution in plasmon-assisted photocatalysis}

\begin{document}

\author{Ieng Wai Un$^{1\ast}$, Yonatan Sivan$^{1,2}$ \\
\normalsize{$^{1}$School of Electrical and Computer Engineering, Ben-Gurion University of the Negev, Israel}\\
\normalsize{$^{2}$Ilse Katz Center for Nanoscale Science and Technology, Ben-Gurion University, Israel}\\
\normalsize{$^\ast$E-mail: iengwai@post.bgu.ac.il}}

\maketitle
%\date{\today}
\begin{abstract}
Recently, there has been a growing interest in the usage of mm-scale composites of plasmonic nanoparticles for enhancing the rates of chemical reactions; the effect was shown recently to be predominantly associated with the elevated temperature caused by illumination. Here, we study the dependence of the temperature distribution on the various parameters of these samples, and provide analytic expressions for simple cases. We show that since these systems are usually designed to absorb all the incoming light, the temperature distribution in them is weakly-dependent on the illumination spectrum, pulse duration, particle shape, size and density. Thus, changes in these parameters yield at most modest quantitative changes. We also show that the temperature distribution is linearly dependent on the beam radius and the thermal conductivity of the host. Finally, we study the sensitivity of the reaction rate to these parameters as a function of the activation energy and show how it manifests itself in various previous experimental reports. These results would simplify the optimization of photocatalysis experiments, as well as of other energy-related applications based on light harvesting for heat generation.
\end{abstract}

\newpage
\section{Introduction}\label{sec:intro}

Metal nanoparticles (NPs) have been studied extensively during the last few decades because of their ability to confine and enhance the electromagnetic field to a sub-wavelength scale. They have found a wide variety of applications, as detailed in some recent reviews~\cite{plasmonics_review_Brongersma_2010,Giannini_chemrev}. In addition, metal NPs have been shown to be ideal heating nanosources when subjected to illumination at their plasmonic resonance wavelength, a research field which is usually referred to as thermo-plasmonics~\cite{Govorov_thermoplasmonics,thermo-plasmonics-basics,thermo-plasmonics-review}. It has led to a wide range of emerging applications such as photo-thermal imaging~\cite{PA_imaging_Orrit,Cognet-phtthm-meth-AnalyChem,Shaked-PT_imaging}, photothermal therapy~\cite{Shaked-PT_imaging,MBCortie-AuNP-hyperthermia-2018,Riley-AuNP-PTT-review-2017,Vines-AuNP-PTT-review-2019}, plasmonic-heating-induced nanofabrication~\cite{Hashimoto-plasmon-heat-fab-glass-2016,Hashimoto-plasmon-nanofab-2016}, and especially those relevant for high temperatures~\cite{refractory_plasmonics,Baffou_solvothermal} and energy applications such as thermo-photovoltaics~\cite{Zubin_thermophotovoltaics,thermal_emission_vlad}, steam generation for purification~\cite{Halas-bubble1,Halas-bubble2,solar_steam_apps,solar_steam_apps_2} and plasmon-assisted photocatalysis~\cite{plasmonic-chemistry-Baffou}. The latter class of experiments was shown in~\cite{Dubi-Sivan-Faraday,anti-Halas-Science-paper,Y2-eppur-si-riscalda,thm_hot_e_faraday_discuss_2019,dyn_hot_e_faraday_discuss_2019,Liu-Everitt-Nano-Letters-2019,Baffou-Quidant-Baldi,Dubi-Sivan-APL-Perspective} to be frequently (even if not always, see~\cite{Baldi-ACS-Nano-2018,yu2018plasmonic,yu2019plasmonic,Boltasseva_LPR_2020,Yugang_Sun,Peng_Chen}) driven by the elevated temperatures that ensue from absorption of light in the metal NPs.

Due to the limited availability of high resolution thermometry techniques (see e.g., discussion in~\cite{Baffou_thermal_imaging,Baffou2014,Baffou-Quidant-Baldi}), efforts were dedicated to modelling the temperature distributions in the samples. Initial studies were dedicated to the characterization of the temperature rise near single nanoparticles under pulsed and continuous wave (CW) illumination~\cite{thermo-plasmonics-basics,thermo-plasmonics-review,Baffou_pulsed_heat_eq_with_Kapitza,Un-Sivan-size-thermal-effect,Qin-ChemSocRev,MinQiu-ACSNano}, including at high intensities~\cite{plasmonic-SAX-PRL,plasmonic-SAX-ACS_phot,plasmonic-SAX-OE,plasmonic-SAX-rods-Ag,japanese_size_reduction,Sivan-Chu-high-T-nl-plasmonics,Gurwich-Sivan-CW-nlty-metal_NP,IWU-Sivan-CW-nlty-metal_NP}. These studies pointed out the importance of the plasmon resonance, and the local nature of the heat generation from the nanoparticles (as opposed to the (nearly uniform) temperature distribution ensuing from macroscopic external heat sources). {\XYZ Some studies also showed how the heating efficiency varies with the NP size~\cite{Un-Sivan-size-thermal-effect} and quantified the relevant temporal and spatial dynamics of the temperature~\cite{thermo-plasmonics-basics,thermo-plasmonics-review,Baffou_pulsed_heat_eq_with_Kapitza,ICFO_Sivan_metal_diffusion,Sivan_Spector_metal_diffusion} in general.} Overall, significant heating of a single particle required relatively intense beams/pulses which are not accessible for many potential applications. 

More recent studies initiated a characterization of the collective thermal response arising in the presence of a large number of NPs~\cite{thermo-plasmonics-multi_NP-Govorov,thermo-plasmonics-multi_NP,Y2-eppur-si-riscalda,Baldi-ACS-Nano-2018,solar_steam_apps,solar_steam_apps_2,Boltasseva_LPR_2020}. As it turns out, the physical picture emerging from these studies is quite different from the one that emerged from the initial single particle studies. In particular, the heat that is initially generated locally at the NPs eventually diffuses into the host and establishes a steady-state temperature distribution in which the total heat generation and the heat loss from the sample to the environment balance each other. In that sense, the difference to the temperature distribution established by an external source may be typically small. This behaviour contrasts the perception of metal NPs as highly localized heat sources on the nano-scale.

In this Article, we re-enforce this view of the temperature distribution emanating from illuminated metal NP ensembles and deepen the related physical insights of such systems. Specifically, as a generic example, we calculate the temperature distribution in a typical sample used in plasmon-assisted photocatalysis experiments. We discuss the sensitivity of the temperature rise and its gradient to various parameters. In particular, we show that for the optically-thick samples which are typical for light-harvesting applications for heating purposes, the temperature rise can be significant even for low illumination levels, and that severe temperature gradients can develop within the samples. We also show that under these conditions, differences related with particle size, shape and density all make, at most, modest quantitative changes. We further show that thermal effects are expected to provide a shallow spectral dependence of reaction rate enhancement, except for cases in which the sample is optically thin and/or the activation energy is high. Finally, we show that the steady-state temperature distribution is determined by the average illumination intensity, such that the temporal pattern, being CW or pulsed, does not affect that distribution.

These results show that claims about the great importance of any of these parameters, or about differences between configurations (e.g., involving gas or liquid hosts, on- or off-resonance illumination, pulsed or CW illumination, large or small NPs etc.) being responsible for qualitative changes in the temperature distribution and reaction rate should be taken with a grain of salt, and better re-examined using quantitative thermal calculations such as those described in the current manuscript. An exception is the sensitivity to the thermal conductivity of the host, which is inversely linear, as well as to the beam radius, which is essentially linear. 

For the specific application of plasmon-assisted photocatalysis, our study is an important step towards a better understanding of thermal effects in conventional photocatalytic experiments especially when re-evaluating thermal effects in previous studies that argued for the dominance of non-thermal effects~\cite{Dubi-Sivan-Faraday,Khurgin-Faraday-hot-es,Y2-eppur-si-riscalda,anti-Halas-Science-paper,R2R,Baffou-Quidant-Baldi,Dubi-Sivan-APL-Perspective}. On the more general level, our work would also be instrumental in uprooting some common misconceptions associated with the role of thermal effects in light harvesting applications that rely on heat generation from a large number of particles. % or from absorbing media in general~\cite{perovskite_solar_cells_comment}.

The paper is organized as follows. We first describe the configuration and the basic assumptions of the model, and develop the model equations for the temperature rise. We then proceed by showing the generic temperature rise distribution and its sensitivity to the various system parameters. Then, we provide a discussion of the results and a comparison to previous experimental works. Finally, we conclude the paper with a brief outlook.

\section{Model and formalism}\label{sec:model}
%Figure~\ref{fig:schematics} shows a prototypical configuration used for plasmon-enhanced chemical reactions. The samples usually consist of a large number ($\sim 10^{12} - 10^{14}$) of metal NPs randomly distributed within a porous metal oxide of a few mm in size.

{\XYZ The typical samples used for plasmon-assisted photocatalysis usually consist of a large number ($\sim 10^{12} - 10^{14}$) of few nm metal NPs randomly distributed within a disc-shaped porous metal oxide of a few mm in size.} % The size of the NPs used in this context is usually no more than a few nm. .. since larger NPs cause strong scattering of photons out of the catalyst sample.} 
Importantly, such systems are usually designed to be optically-thick such that all the illumination energy is absorbed; the reaction rate is then enhanced due to the elevated temperature~\cite{Y2-eppur-si-riscalda}.

\begin{figure}[h]
\centering
\includegraphics[width=1\textwidth]{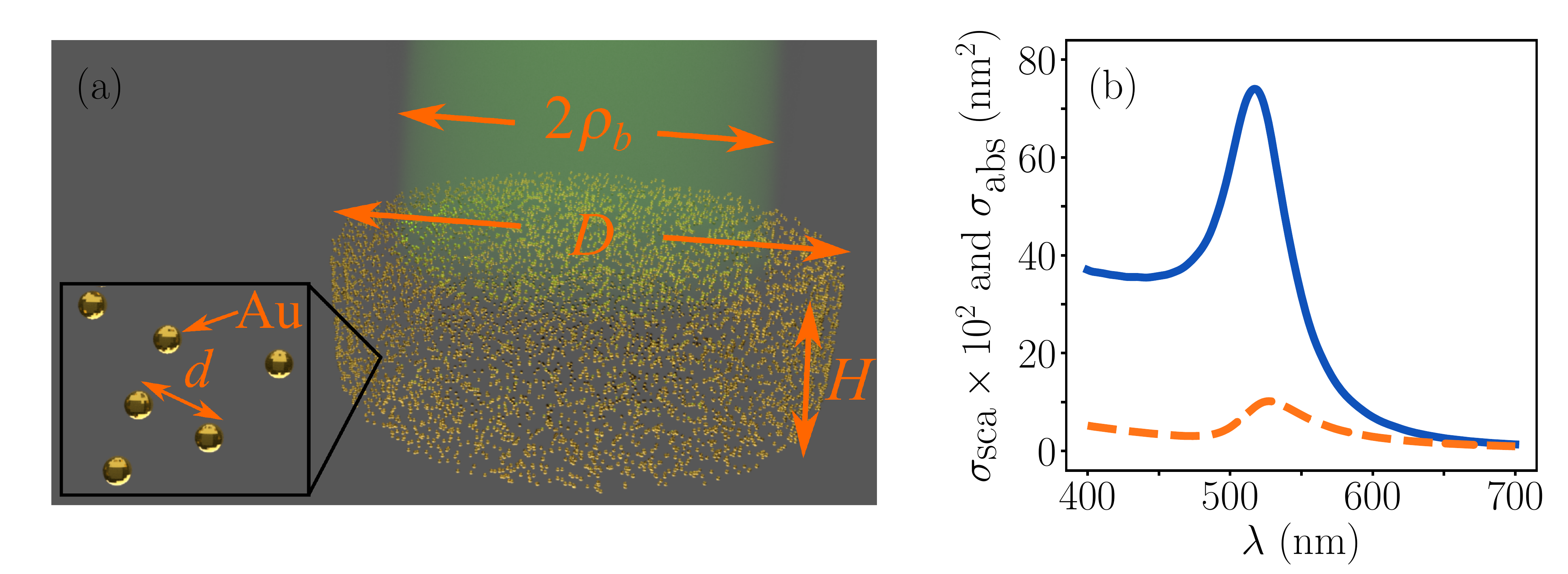}
\caption{(Color online) (a) {\XYZ A schematic of an illuminated many-nanoparticle system in which many identical spherical metal NPs with radius $a$ are randomly distributed in a disc-shape (with $D$ being the diameter and $H$ being the thickness). $d$ is the (average) inter-particle spacing between NPs and $\rho_b$ is the beam radius of the illumination. (b) The wavelength-dependence of the absorption cross-section (blue solid line) and of the scattering cross-section (orange dashed line) of a single 6 nm radius Au NP immersed in porous oxide host. } }\label{fig:schematics}
\end{figure}

{\XYZ A realistic model of samples used for photocatalysis has to take into account the thermal properties (in particular, the thermal conductivity) of the various components of the system - the catalyst, the solid chamber, the inner host (gas or liquid) and outer host (usually air), the window through which the illumination enters the sample etc.. This can be done only using high-end numerical methods, like finite element methods, to achieve accurate temperature distribution of the system. However, not only such simulations require expertise, computational resources and are time-consuming, they also do not readily provide simple insights about the underlying physics. In particular, this includes general trends in the dependence of the temperature distribution on the various parameters of many-nanoparticle system and a better understanding of the relationship between the macroscopic temperature rise and the light absorption on the nanoscale.

Thus, in order to obtain such insights, we simplify the system by assuming that the sample consists of {\em identical spherical} metal NPs with radius $a$ (with dielectric permittivity $\varepsilon_m = \varepsilon_m^{\prime} + i\varepsilon_m^{\prime\prime}$ and thermal conductivity $\kappa_m$) distributed in a disc shape and immersed in a uniform host material with $d$ being the average inter-particle spacing of the randomly distributed NPs, as shown in Figure~\ref{fig:schematics}(a).} The (effective) permittivity and thermal conductivity of the host material ($\varepsilon_h$ and $\kappa_h$, respectively) are related to the volume fraction of air and oxide via the Maxwell Garnett equation~\cite{maxwell1881treatise,birdtransport,Pietrak-eff-kappa-compos-2014,Y2-eppur-si-riscalda}. %These approximations are good for pellet geometries, liquid suspensions etc. alike and, in fact, unavoidable. 
As we show in Supplemental Information Section~\ref{app:comsol_non_uniform_kappa} and~\ref{app:comsol_liquid_kappa} , all the trends we identify in the consequent simulations are of a general nature, i.e., they are valid also when a realistic configuration is studied using exact numerical simulations.

The NPs are heated by either a CW or a pulsed illumination with a beam spot radius of $\rho_b$. For monochromatic CW illumination (pulsed illumination), we denote the angular frequency (central angular frequency) by $\omega = 2\pi c/\lambda$, where $\lambda$ is the wavelength and $c$ is the speed of light in vacuum; the illumination intensity (time average illumination intensity) of the laser is denoted by $I_{\textrm{inc}}$ ($\langle I_{\textrm{inc}} \rangle$). 

Under monochromatic CW illumination of low intensity, the temperature distribution can be obtained by properly summing the heat generated by all NPs in the system~\cite{Y2-eppur-si-riscalda} (see details in Supplementary Information Section~\ref{app:temp_distrib_CW})
\begin{align}\label{eq:mti_NP_sum}
\Delta T(\omega,{\bf r}) = \begin{cases}
\dfrac{I_{\textrm{inc}}\sigma_{\textrm{abs}}(\omega)}{4 \pi \kappa_h}\left[\dfrac{e^{-z_i/\delta_{\textrm{skin}}(\omega)}}{a} + \displaystyle\sum\limits_{j\neq i}\dfrac{e^{-z_j/\delta_{\textrm{skin}}(\omega)}}{|{\bf r}_j - {\bf r}_i|}\right],&\textrm{for NP at } {\bf r}_i,\\
\hfil\dfrac{I_{\textrm{inc}}\sigma_{\textrm{abs}}(\omega)}{4 \pi \kappa_h }\displaystyle\sum\limits_{j} \dfrac{e^{-z_j/\delta_{\textrm{skin}}(\omega)}}{|{\bf r}_j - {\bf r}|},& \textrm{ for } {\bf r} \textrm{ in the host.}
\end{cases}
\end{align}
Here, $z_i$ is the $z$-coordinate of the position of $i$-th NP, $\sigma_{\textrm{abs}}$ is the absorption cross-section of the NP, and $\delta_{\textrm{skin}}$ is the skin (penetration) depth (equivalently, the inverse of the absorption coefficient) experienced by the incident beam; it can be approximated by the NP density and absorption cross-section as~\cite{Y2-eppur-si-riscalda} 
\begin{align}\label{eq:delta_skin}
\delta_{\textrm{skin}}(\omega) = d^3/\sigma_{\textrm{abs}}(\omega).
\end{align}
It has been shown~\cite{thermo-plasmonics-multi_NP} that two distinct regimes of the temperature profile can be achieved: a temperature confinement regime where the temperature rise is confined at the vicinity of each NP, and a temperature delocalization regime where the temperature profile is smooth throughout the composite. The former regime is realized only when a small number of NPs ($< 10^3$) is illuminated~\cite{thermo-plasmonics-multi_NP} (either because the NP density is highly dilute or because the beam size is small). {\XYZ In these cases, the overall temperature rise is larger when the particle density increases.} These configurations might be useful from the physical perspective (e.g., when attempting to identify the origin of chemical reactions~\cite{Boltasseva_LPR_2020}), but are, however, of little practical relevance because they enable only limited heating\footnote[1]{If the activation energy $E_a$ is low, then, such systems may benefit from non-thermal ``hot'' electron action, see~\cite{Boltasseva_LPR_2020}.}. However, in applications of plasmon-assisted photocatalysis, the beam is typically wide enough and there are usually many more NPs under the illumination. In this case, the overall temperature rise of each NP is dominated by the contribution from other NPs, as we shall see later. As a result, the temperature profile is almost completely smooth throughout the sample.

\section{Results}\label{sec:results}
\subsection{CW illumination}\label{sec:cw_illumination}
{\XYZ We consider a configuration which consisted of a Au NP ensemble immersed in a host material with $\varepsilon_h = 1.44$ and $\kappa_h = 50.3$ mW/(m$\cdot$K) corresponding to porous oxide host.} Initially, we assume that the NP size has a radius of $a = 6$ nm, the (average) inter-particle spacing is $d = 225$ nm, the thickness of the NPs array is $H = 1$ mm, {\XYZ the NPs are illuminated at $\lambda = 532$ nm with $I_{\textrm{inc}} = 80$ mW/cm$^2$ and an} illumination spot area of $\pi \rho_b^2 = 1$ cm$^2$. We calculate the temperature distribution of the system by a numerical solution of Eq.~(\ref{eq:mti_NP_sum}). Then, we compare the numerical solution of Eq.~(\ref{eq:mti_NP_sum}) to its approximation, and also test the validity of our effective medium approximation of the heat source.

\subsubsection{Generic temperature distribution}\label{sec:temp_distrib}

\begin{figure}[h]
\centering
\includegraphics[width=1\textwidth]{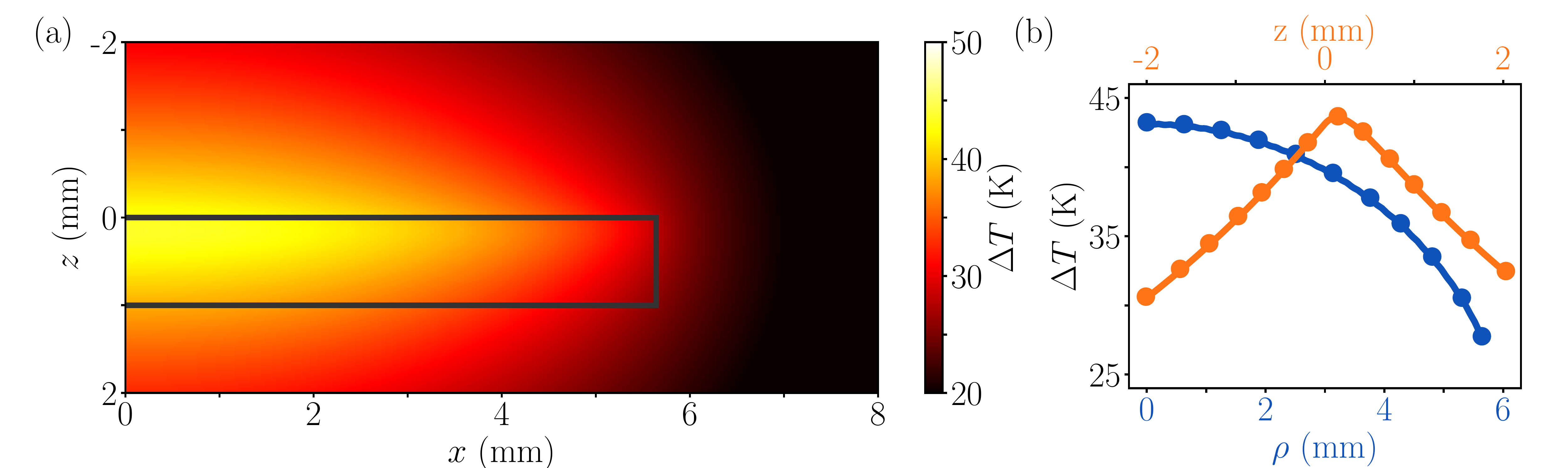}
\caption{(Color online) {\XYZ (a) The temperature rise $\Delta T$ of the photocatalyst sample (the edges of which are represented by the black lines) calculated by Eq.~(\ref{eq:mti_NP_sum}). The photocatalyst sample is illuminated from the top by a CW laser at the wavelength $\lambda = 532$ nm with illumination intensity $I_{\textrm{inc}} = 80$ mW/cm$^2$. The inter-particle spacing is 225 nm and the beam width is the same as the diameter of the photocatalyst sample. (b) The temperature rise profile along the illumination direction (blue solid line) and the temperature rise profile on the surface facing the light source (orange solid line) obtained by Eq.~\eqref{eq:mti_NP_sum}. The dots represent the COMSOL simulation results which are seem to agree perfectly with Eq.~\eqref{eq:mti_NP_sum}.}} \label{fig:dtemp_distrib_rev}
\end{figure}

The absorption and scattering cross-sections of the Au NPs in the sample are calculated by using the permittivity from~\cite{PT_Shen_ellipsometry_gold}, {\XYZ see Figure~\ref{fig:schematics}(b)}. For the small NPs we are considering, one can see that $\sigma_{\textrm{abs}} \gg \sigma_{\textrm{sca}}$. The domination of absorption over scattering justifies a-posteriori the effective medium approximation~\footnote[2]{NPs larger than 20 nm in radius give rise to significant scattering. In this case, if the NP density is relatively low, as in this study, one can model the light intensity in the catalyst sample by using the effective medium approach accounting the multiple scattering ~\cite{sihvola_EM_mixing,choy_EMT,Y2-eppur-si-riscalda} or by the equation of radiative transfer~\cite{tsang_scattering_2004,ishimaru_wave_scattering_random_media_1978,chandrasekhar_radiative_transfer_2013}. Either way, the exponential decay due to the absorption is not significantly modified due to the effects of multiple scattering~\cite{sihvola_EM_mixing,choy_EMT,tsang_scattering_2004}.} (used in Supplementary Information Section~\ref{app:temp_distrib_CW}). For $\lambda = 532$ nm, {\XYZ the penetration (skin) depth is $\delta_{\textrm{skin}} \approx 0.19$ mm,} indeed much shorter than thickness of the NPs array $H$. The results of the calculation of the temperature rise $\Delta T$ using Eq.~(\ref{eq:mti_NP_sum}) are shown in Figure~\ref{fig:dtemp_distrib_rev}(b). {\XYZ We also use COMSOL to simulate the temperature rise distribution in good agreement with the numerical solution of Eq.~\eqref{eq:mti_NP_sum}, see Figure~\ref{fig:dtemp_distrib_rev}(b). One can see that $\Delta T$ along the illumination direction drops from $\sim 43$ K to $\sim 30$ K at a distance of $\sim 2$ mm from the surface, and the temperature rise on the surface facing %(away from) 
the light source decreases gradually from $\sim 43$ K %(40 K) 
at the center to 25 K %(28 K) 
at the edge.} The overall temperature rise is much higher than the temperature rise in the single-particle problem ($\sim$ 10 $\mu$K) given by Eq.~(\ref{eq:heat_eq_sol_sng_NP_CW}). This indicates that the overall temperature rise is, indeed, a many-particle effect. The temperature rise at the center is higher than at the edges because this region benefits from heat arriving from all directions, whereas in the periphery, it arrives only from the center. This temperature nonuniformity shows that a standard normalization of the reaction rates in photocatalysis by the catalyst mass (as e.g., in~\cite{Halas_Science_2018,Halas-Nature-Energy-2020}) can incur severe errors in evaluation of the reaction enhancement rate, see discussion in~\cite{Y2-eppur-si-riscalda,R2R}.

In order to test the sensitivity of the results to the exact particle positions, we compare the temperature distribution for a periodic array and a fully random array of NPs. Specifically, we start with a regular cubic NPs array (all other parameters are left unchanged). Next, we move each NP in the $x$-, $y$- and $z$-directions by a random amount ranging between $-d/2+a$ and $d/2-a$. Then, we sum the contributions from each NP and {\XYZ obtain $\Delta T^{\textrm{top}} = 44.2$ K,} very close to the result of the regular NPs array. This shows that the randomness of the arrangement of a very large number of NPs has an insignificant effect on the overall temperature rise, thus, justifying the effective medium approximation of the heat source.  

The summation over NPs in Eq.~(\ref{eq:mti_NP_sum}) can be approximated by an equivalent integration~\cite{thermo-plasmonics-multi_NP}. As we show in Supplementary Information Section~\ref{app:est_dT1_st_CW}, this equivalent integration enables an approximation of the temperature increase at the center of the top surface (the surface facing the light source, denoted by $\Delta T^{\textrm{top}}$ hereafter) whereby
\begin{align}\label{eq:dTtop_CW}
\Delta T^{\textrm{top}} \approx \dfrac{I_{\textrm{inc}}\rho_0}{2\kappa_h} \left[1 - e^{- H/\delta_{\textrm{skin}}(\omega)}\right].
\end{align}
Here, $\rho_0 = \min(\rho_b,D/2)$ represents the radius of illuminated NPs (see details in Supplementary Information Section~\ref{app:est_dT1_st_CW}) and $H$ is the sample thickness. In experiments, the beam size is typically set to $\rho_b \lesssim D/2$ so that all the illumination energy can be absorbed. In this case, one can simply replace $\rho_0$ by $\rho_b$. {\XYZ For the photocatalyst sample shown in Figure~\ref{fig:schematics} (a), the approximation of $\Delta T^{\textrm{top}}$ given by Eq.~(\ref{eq:dTtop_CW}) is calculated to be 44.6 K, in good agreement with the numerical solution of Eq.~(\ref{eq:mti_NP_sum}).}

\begin{figure}[h]
\centering
\includegraphics[width=0.4\textwidth]{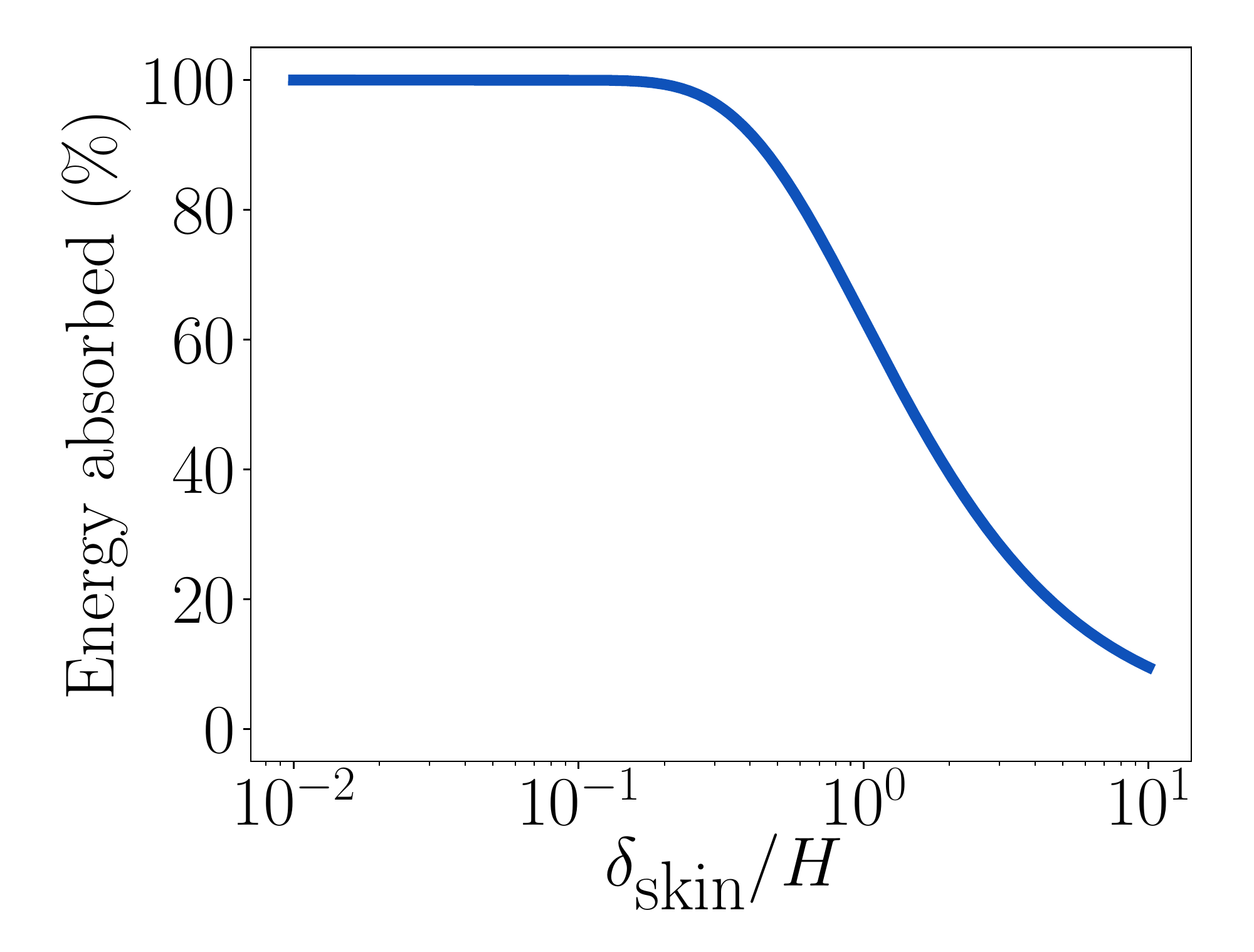}
\caption{(Color online) The energy absorbed (in \%) by the sample as a function of $\delta_{\textrm{skin}}/H$.} \label{fig:abseng_vs_delta_skin}
\end{figure}

The expression~(\ref{eq:dTtop_CW}) indicates that the overall (approximate) temperature increase is proportional to the fraction of the illumination energy absorbed by the sample, which itself is related to the ratio of sample thickness to the penetration (skin) depth ($\sim 1 - \exp(-H/\delta_{\textrm{skin}})$). This is shown in Figure~\ref{fig:abseng_vs_delta_skin} via the dependence of the relative energy absorbed by the sample on the ratio of the penetration (skin) depth and the sample thickness. One can see that more than 99\% of the illumination energy is absorbed if $\delta_{\textrm{skin}}/H < 0.2$. Therefore, for optically thick samples (i.e., $\delta_{\textrm{skin}}\ll H$), the overall temperature increase is expected to be weakly-dependent on $\sigma_{\textrm{abs}}$ and on the inter-particle spacing $d$ (both via $\delta_{\textrm{skin}}$~(\ref{eq:delta_skin})), but to be much more sensitive to the thermal conductivity of the host medium (see detailed discussion below). In that respect, the temperature~(\ref{eq:dTtop_CW}) is essentially the same as for an infinitely thin disc-shaped heat source in free space. Potentially unexpectedly, this also implies that {\em the temperature distributions in thick structures will exhibit relatively weak spectral features}. In contrast, for thin samples (e.g., a monolayer of NPs, e.g., in~\cite{thermo-plasmonics-multi_NP}), Eq.~(\ref{eq:dTtop_CW}) reduces to 
\begin{align}\label{eq:dTtop_CW_thin}
\Delta T^{\textrm{top}} \approx \dfrac{\sigma_{abs} I_{\textrm{inc}}}{2\kappa_h d} \dfrac{\rho_0}{d}.
\end{align}
Thus, in comparison to the heating of a single NP by a plane wave, the heat generation from the additional NPs causes the enhancement of the temperature rise by a factor of $2 \pi \rho_0 a/ d^2$. 

{\XYZ Eq.~\eqref{eq:dTtop_CW} also indicates that there exist two distinct regimes of the wavelength dependence of the temperature rise: a regime where the temperature rise has a similar wavelength dependence to the absorption cross section of the single particle, and a regime where the temperature rise is less sensitive to the illumination wavelength. The former regime occurs when the absorption cross section is very small ($\sigma_\textrm{abs}\ll d^3/H$, e.g. the illumination wavelength is far from the plasmon resonance wavelengths) or/and the particle density is highly dilute ($d^3\gg H\sigma_\textrm{abs}$). In these case, Eq.~\eqref{eq:dTtop_CW} reduces to 
\begin{align}
\Delta T^\textrm{top} \approx \dfrac{\sigma_\textrm{abs} I_\textrm{inc}}{2\kappa_h}\dfrac{\rho_0 H}{d^3},
\end{align}
and the temperature rise is proportional to the absorption cross-section, i.e., they have the same spectral dependence. However, as mentioned in Section~\ref{sec:model}, these configuration are of less relevance to applications of plasmon-assisted photocatalysis or light harvesting. }

In what follows, we study the sensitivity of the temperature profile to the various parameters of the system.

\subsubsection{Inter-particle-spacing- and NP-size-dependence of the temperature distribution}\label{sec:a-d-dep-DeltaT}

\begin{figure}[h]
\centering
\includegraphics[width=1\textwidth]{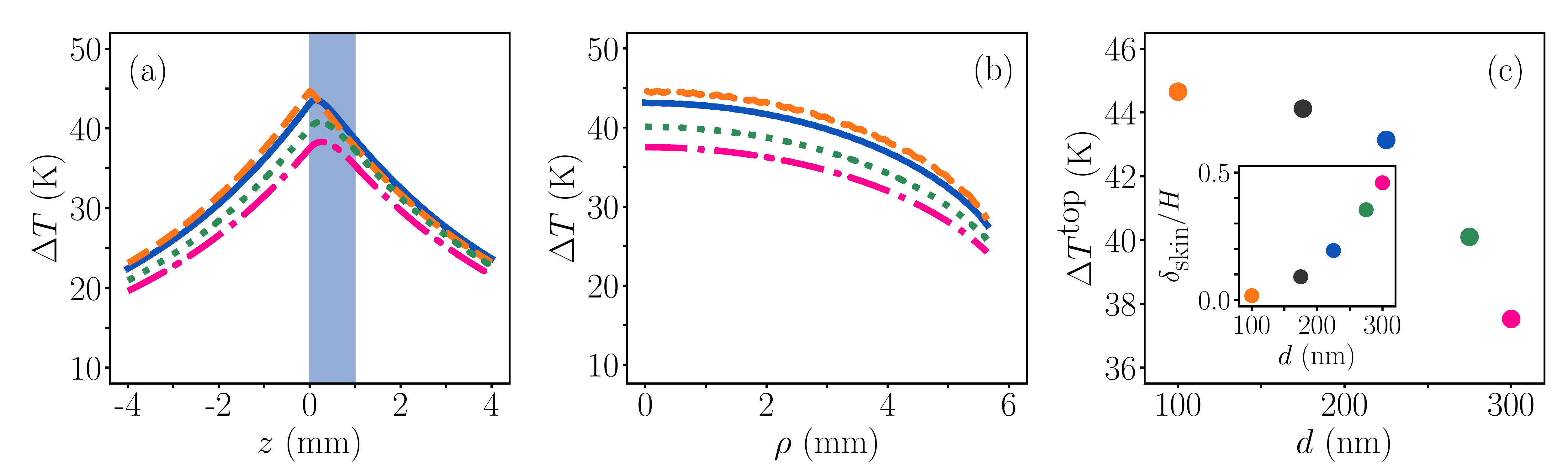}
\caption{(Color online) {\XYZ (a) The temperature rise profile along the illumination direction and (b) the temperature rise profile on the surface facing the light source for} {\XYZ inter-particle spacing $d =$ 100 nm (orange dashed line), 225 nm (blue solid line), 275 nm (green dotted line) and 300 nm (magenta dash-dotted line).} The blue-gray region represents the NPs array. All other parameters are the same as %Figure~\ref{fig:dtemp_distrib_rev}(b). 
in Section~\ref{sec:temp_distrib}.
(c) The temperature rise at the center of the top surface for different inter-particle spacing. The inset shows the corresponding inter-particle-spacing-dependence of $\delta_{\textrm{skin}}/H$. The colored symbols correspond to the $\Delta T$ profile marked by the same color in (a) and (b). }\label{fig:dtemp_vs_d_au6} %Figure~\ref{fig:cabs_csca_dtemp_au_6nm_H2O_532nm}.}\label{fig:dtemp_vs_d_au6}
\end{figure}

In Figure~\ref{fig:dtemp_vs_d_au6}(a) and (b) we plot the temperature rise profile along the illumination direction and on the surface facing the light source for different inter-particle spacing with all other parameters being the same as in Sec.~\ref{sec:temp_distrib}. One can see that the overall temperature rise is insensitive to the inter-particle spacing. Indeed, {\XYZ the overall $\Delta T$ decreases by $\sim$ 1.6 K ($< 5$\%) when $d$ changes from 100 nm to 225 nm (NP density decreases by 90\%), whereas it decreases by $\sim$ 6 K ($\sim 13$\%) when $d$ further changes from 225 nm to 300 nm (NP density decreases by 60\%)}. As discussed above, the {\it weak} dependence of $\Delta T$ on the inter-particle spacing can be understood by the fact that the overall temperature rise depends primarily on the amount of photon energy absorbed by the sample. In order to demonstrate that explicitly, Figure~\ref{fig:dtemp_vs_d_au6}(c) shows the $d$-dependence of $\Delta T$ at the center of the top surface of the sample, with the corresponding $\delta_{\textrm{skin}}/H$ shown in the inset. {\XYZ One can see that when $d < 225$ nm, $\delta_{\textrm{skin}}/H < 0.2$ such that more than 99\% of the photon energy is absorbed (see Figure~\ref{fig:abseng_vs_delta_skin}); when 225 nm $<d<$ 300 nm, $\delta_{\textrm{skin}}/H$ increases from 0.2 to 0.5 and the absorbed photon energy decreases from 99\% to 86\% (see Figure~\ref{fig:abseng_vs_delta_skin}).} This not only explains the weak dependence of the overall temperature rise $\Delta T$ on the inter-particle spacing when the skin depth is smaller than the sample thickness, {\XYZ but also shows that the $d$-dependence of the overall $\Delta T$ is even weaker when $d < $ 225 nm.} In that respect, the $d$-dependence of $\Delta T$ for the current 3D arrangement of NPs is {\it much weaker} than that of a single-layer array~\cite{thermo-plasmonics-multi_NP}.

{\XYZ In passing, it is worth noting that the maximal temperature does not necessarily occur at the center of the surface facing the light source, especially when the skin depth is compatible to the sample thickness, see the case of $d = 300$ nm in Figure~\ref{fig:dtemp_vs_d_au6}(a).} In this case, the heating source becomes more uniformly distributed in the sample so that the maximal temperature occurs inside the sample. {\XYZ The maximum temperature occuring inside the sample and the temperature gradient imposes difficulty on the use of thermal cameras for temperature determination in experiments} (see more discussions in Refs.~\cite{anti-Halas-Science-paper,Y2-eppur-si-riscalda,R2R}).

\begin{figure}[h]
\centering
\includegraphics[width=0.7\textwidth]{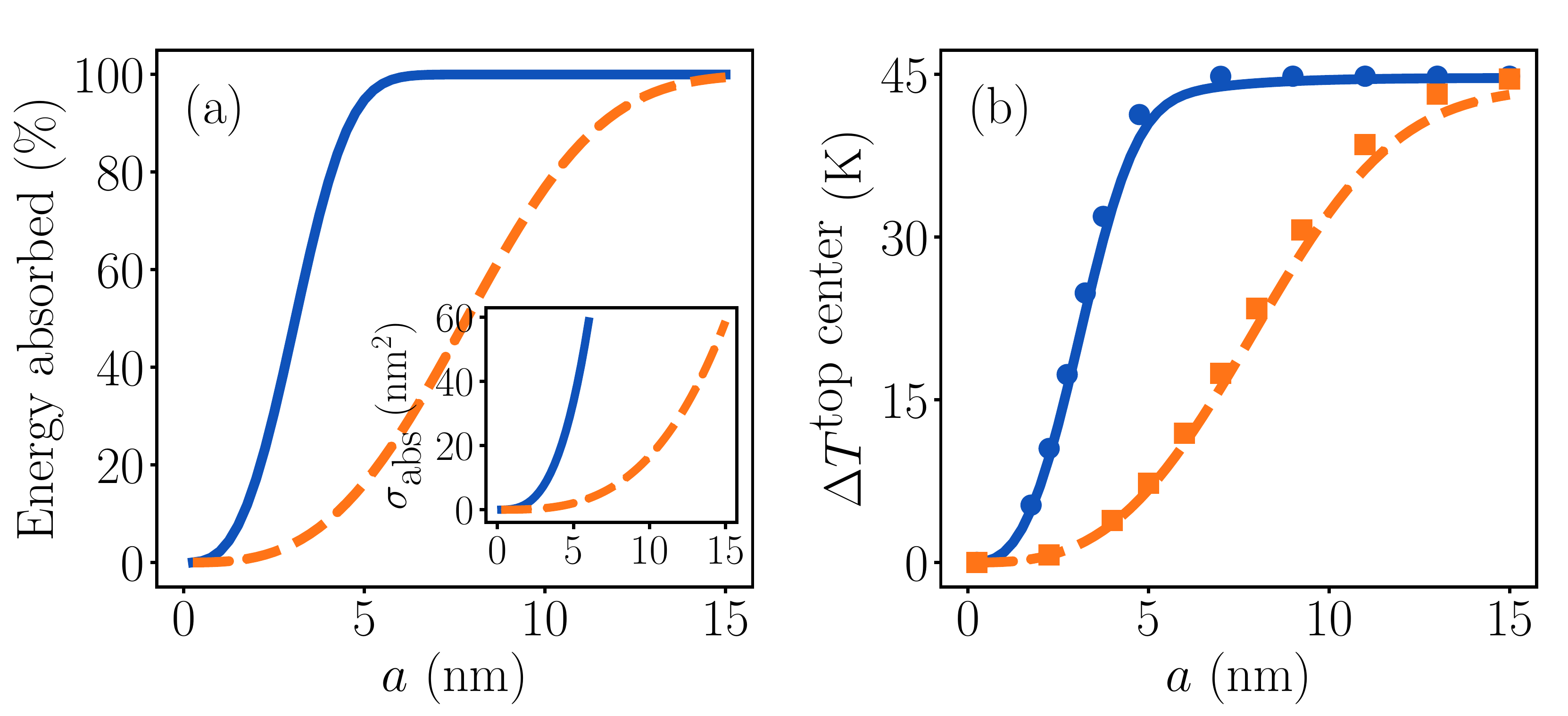}
\caption{(Color online) {\XYZ (a) The illumination energy energy absorbed by the sample (in \%) as a function of the particle radius $a$ (inset: the particle radius dependence of the absorption cross section). (b) $\Delta T^\textrm{top}$ as a function of the particle radius $a$. In (a) and (b), the blue solid lines and the orange dashed lines represent for the wavelength $\lambda = 532$ nm and $\lambda = 633$ nm, respectively. The blue circles and the orange squares in (b) represent the $\Delta T^\textrm{top}$ obtained from the approximate analytical solution Eq.~(\ref{eq:dTtop_CW}) for $\lambda = 532$ nm and $\lambda = 633$ nm, respectively.} All other parameters are the same as %Figure~\ref{fig:dtemp_distrib_rev}(b). %
in Section~\ref{sec:temp_distrib}.
}\label{fig:dtemp_vs_a_au_wl}%Figure~\ref{fig:cabs_csca_dtemp_au_6nm_H2O_532nm}.}\label{fig:dtemp_vs_a_au_wl}
\end{figure}

In a similar manner, we now study the size-dependence of the overall temperature rise by calculating the energy absorbed by the sample and $\Delta T^{\textrm{top}}$ for the photocatalyst sample shown in Figure~\ref{fig:dtemp_distrib_rev}(b) with different sizes of Au NPs, see Figure~\ref{fig:dtemp_vs_a_au_wl}. The wavelength of the illumination is chosen to be either within the plasmon resonance bandwidth ($\lambda = $ 532 nm) or out of the plasmon resonance bandwidth ($\lambda = $ 633 nm). Apart from the particle size and the wavelength, all other parameters are the same in Figure~\ref{fig:dtemp_distrib_rev}(b). One can see that for $\lambda = $ 532 nm ($\lambda = $ 633 nm), {\XYZ the overall $\Delta T$ depends weakly on the particle size when $a > 5$ nm ($a > 15$ nm) but it decreases strongly with decreasing particle size when $a < 5$ nm ($a < 15$ nm).} Moreover, the NP-size-dependence of $\Delta T^{\textrm{top}}$ can be again explained by the relation between absorbed energy and the ratio of the skin depth to the sample thickness via Eq.~(\ref{eq:delta_skin}). Since the absorption cross-section increases (and the skin depth decreases) with the NP size (see the inset in Figure~\ref{fig:dtemp_vs_a_au_wl} (a))~\cite{Bohren-Huffman-book,Un-Sivan-size-thermal-effect}, once the NP size is large enough such that $\delta_{\textrm{skin}}/H < 0.3$, more than 95\% of the illumination energy is absorbed by the sample and the overall $\Delta T$ shows a weak dependence on the NP size. 

\subsubsection{Wavelength-dependence of the temperature distribution}\label{sec:wavelength_dep_T}
\begin{figure}[h]
\centering
\includegraphics[width=0.7\textwidth]{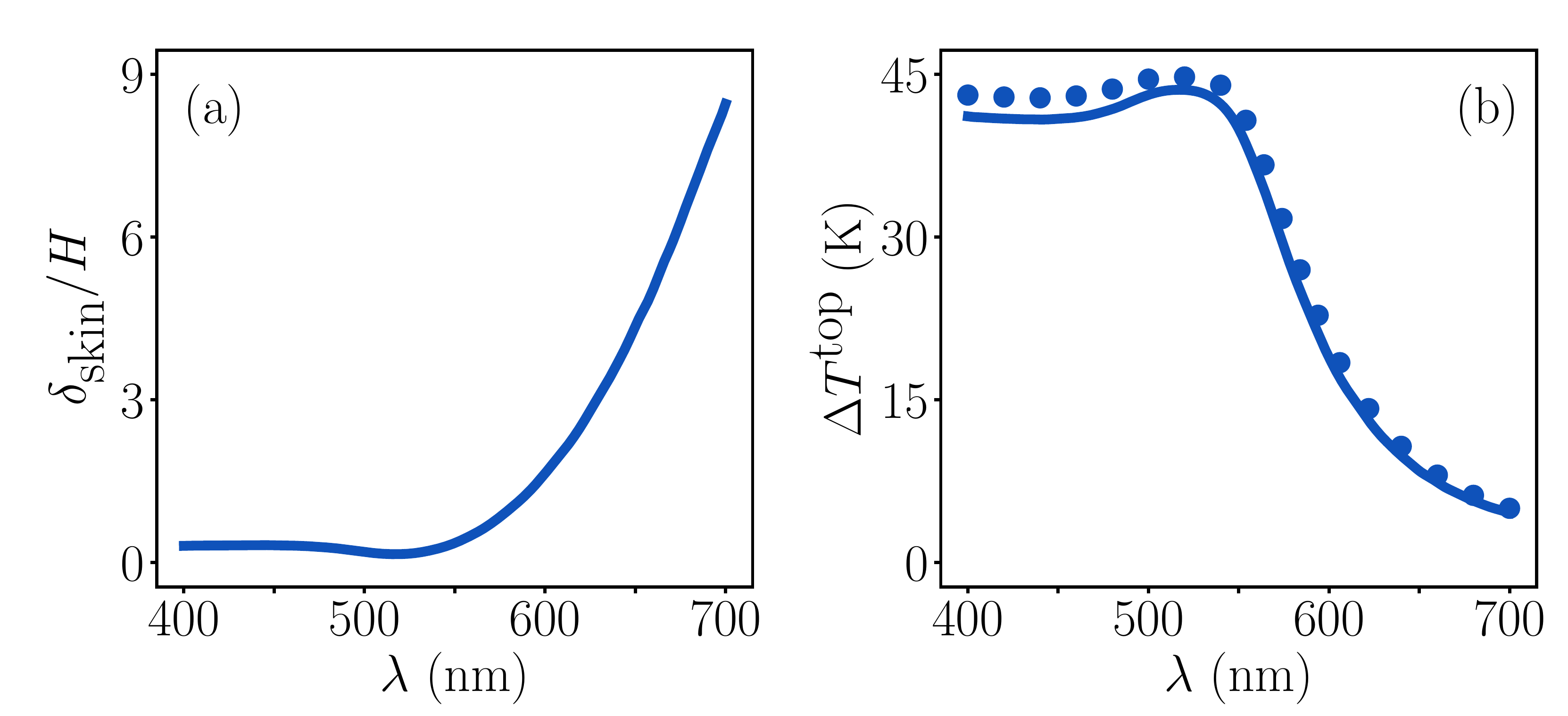}
\caption{(Color online) (a) {\XYZ The spectral dependence of the ratio of the skin depth to the sample thickness.} (b) The temperature rise Eq.~(\ref{eq:mti_NP_sum}) (blue solid line) and its approximate analytical solution Eq.~(\ref{eq:dTtop_CW}) (blue dots) at the center of the top surface of the NP array as a function of illumination wavelength. All other parameters are the same as %Figure~\ref{fig:dtemp_distrib_rev}(b).%
in Section~\ref{sec:temp_distrib}.
}\label{fig:dtemp_vs_wl_au6}
\end{figure}

To study the wavelength-dependence of the overall temperature rise, we now assume that the photocatalyst sample shown in Figure~\ref{fig:dtemp_distrib_rev}(b) is illuminated by a ``tunable'' single-wavelength CW source~\cite{Halas_dissociation_H2_TiO2,Halas_H2_dissociation_SiO2,Halas_Science_2018} with {\XYZ a fixed illumination intensity of 80 mW/cm$^2$}, and calculate the penetration depth (\ref{eq:delta_skin}) and $\Delta T^{\textrm{top}}$ using Eq.~(\ref{eq:mti_NP_sum}) as a function of the illumination wavelength, see Figure~\ref{fig:dtemp_vs_wl_au6}. {\XYZ For 400 nm $<\lambda<$ 560 nm, $\delta_{\textrm{skin}}/H < 0.5$}, more than 86\% of the illumination energy is absorbed by the sample (see Figure~\ref{fig:abseng_vs_delta_skin}) such that the overall temperature rise exhibits a fairly weak $\lambda$-dependence, much weaker than that of the absorption cross-section as shown in Figure~\ref{fig:dtemp_distrib_rev} (a). The $\Delta T^{\textrm{top}}$ at the plasmonic resonance wavelength is only 4\% higher than the short wavelength shoulder. However, {\XYZ for $\lambda > 610$ nm, $\delta_{\textrm{skin}}/H > 2$}, so that the illumination energy absorbed by the sample and the overall $\Delta T$ are roughly proportional to $H/\delta_{\textrm{skin}}$, thus roughly proportional to $\sigma_{\textrm{abs}}$. {\XYZ This wavelength-dependent behavior is strictly different from the case of single particle system in which the temperature rise is proportional to the (spectrally narrow) absorption cross-section.} The wavelength-dependent $\Delta T^{\textrm{top}}$ obtained by the approximate analytical solution Eq.~(\ref{eq:dTtop_CW}) is again in excellent agreement with the exact numerical solution of Eq.~(\ref{eq:mti_NP_sum}). 

We note that the energy absorbed by the sample and the overall temperature rise will exhibit a similar dependence on the absorption cross-section (e.g., by variation of the host or NP permittivity) as their dependence on the particle size and the wavelength (Figs.~\ref{fig:dtemp_vs_a_au_wl} and~\ref{fig:dtemp_vs_wl_au6}, respectively). Accordingly, such specific simulations are not shown.

\subsubsection{Beam width-dependence of the temperature distribution}
The dependence of the overall temperature rise on the illumination beam radius can be separated into two distinct regimes, depending on the relative size of the beam spot with respect to the sample surface. In each regime, one can vary the beam size with either the illumination intensity or the illumination power being fixed.

\begin{figure}[h]
\centering \includegraphics[width=1\textwidth]{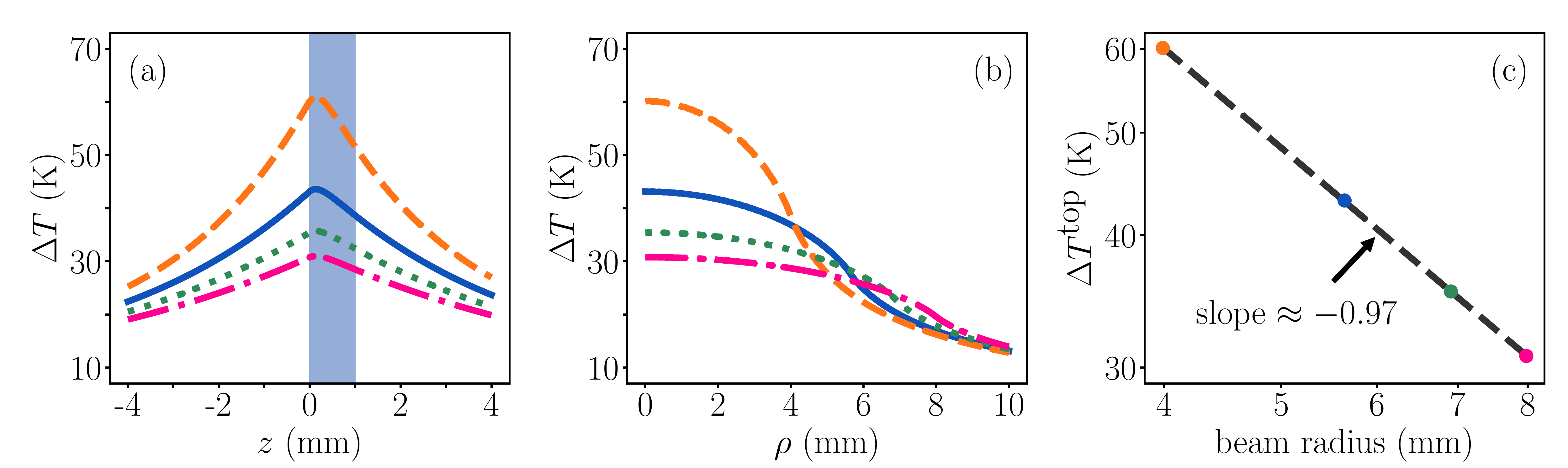}
\caption{(Color online) {\XYZ (a) The temperature rise profile along the illumination direction and (b) the temperature rise profile at the surface facing the light source} for beam spot area 0.5 cm$^2$ (orange dashed line), 1 cm$^2$ (blue solid line), 1.5 cm$^2$ (green dotted line) and 2 cm$^2$ (magenta dash-dotted line) when the illumination power is fixed to be 1 W. The blue-gray region represents the NPs array. All other parameters are the same as %Figure~\ref{fig:dtemp_distrib_rev}(b). %
in Section~\ref{sec:temp_distrib}. %Figure~\ref{fig:cabs_csca_dtemp_au_6nm_H2O_532nm}.
(c) $\Delta T^{\textrm{top}}$ as a function of the beam radius (log-log scale). The colored symbols correspond to the $\Delta T$ profile marked by the same color in (a) and (b). The slope $\approx$ -1 indicates that $\Delta T^{\textrm{top}} \propto \rho_b^{-1}$. }\label{fig:dtemp_vs_bmsz_au6}
\end{figure}

When the beam radius is smaller than the sample radius, i.e., when $\rho_b < D/2$, the number of the NPs under illumination is proportional to the beam spot area such that we set $\rho_0 = \rho_b$ in Eq.~(\ref{eq:dTtop_CW}). If the illumination power is fixed, then, when the beam radius decreases, the illumination intensity increases but the total illumination energy remains the same, so that the overall temperature rise is higher, see Figure~\ref{fig:dtemp_vs_bmsz_au6}(a)-(b). If we plot $\Delta T^{\textrm{top}}$ obtained by Eq.~(\ref{eq:mti_NP_sum}) as a function of the beam radius in log-log scale (see Figure~\ref{fig:dtemp_vs_bmsz_au6}(c)), {\XYZ one can see that $\Delta T^{\textrm{top}}$ is roughly inversely proportional to the beam radius (slope $\approx -0.97$)}. This is found to be in excellent agreement with the result deduced from Eq.~(\ref{eq:dTtop_CW}) that $\Delta T^{\textrm{top}} \propto I_{\textrm{inc}} \rho_b \propto \rho_b^{-1}$. If the illumination intensity is fixed, $\Delta T$ is thus proportional to the beam radius $\rho_b$ (not shown). 

When the beam radius is larger than the sample radius, i.e., $\rho_b > D/2$, we set $\rho_0 = D/2$ in Eq.~(\ref{eq:dTtop_CW}). If the illumination power is fixed, the illumination intensity experienced by the NPs $I(\omega,{\bf r}_i)$ and thus $\bar{p}_{\textrm{abs},i}$ increase linearly with decreasing the beam spot area ($\bar{p}_{\textrm{abs},i} \propto I(\omega,{\bf r}_i) \propto \rho_b^{-2}$), but the number of NPs under illumination remains unchanged. As a result, it can be shown from Eq.~(\ref{eq:mti_NP_sum}) and (\ref{eq:dTtop_CW}) that the overall $\Delta T$ is inversely proportional to the beam spot area, namely, $\Delta T \sim \rho_b^{-2}$ (not shown). If the illumination intensity is fixed, $\Delta T$ is independent of the beam radius (not shown). 

Notably, the different scaling of the temperature rise with the beam size also differ from the scaling of non-thermal effects with the beam size~\cite{Baffou-Quidant-Baldi}. Accordingly, Baffou {\em et al.} recently suggested in~\cite{Baffou-Quidant-Baldi} that these behaviours can be used to separate the contributions of thermal and non-thermal effects in plasmon-assisted catalysis reactions.

\subsubsection{Thermal conductivity-dependence of the temperature distribution}\label{sec:kappa_dep_T}
\begin{figure}[h]
\centering \includegraphics[width=0.7\textwidth]{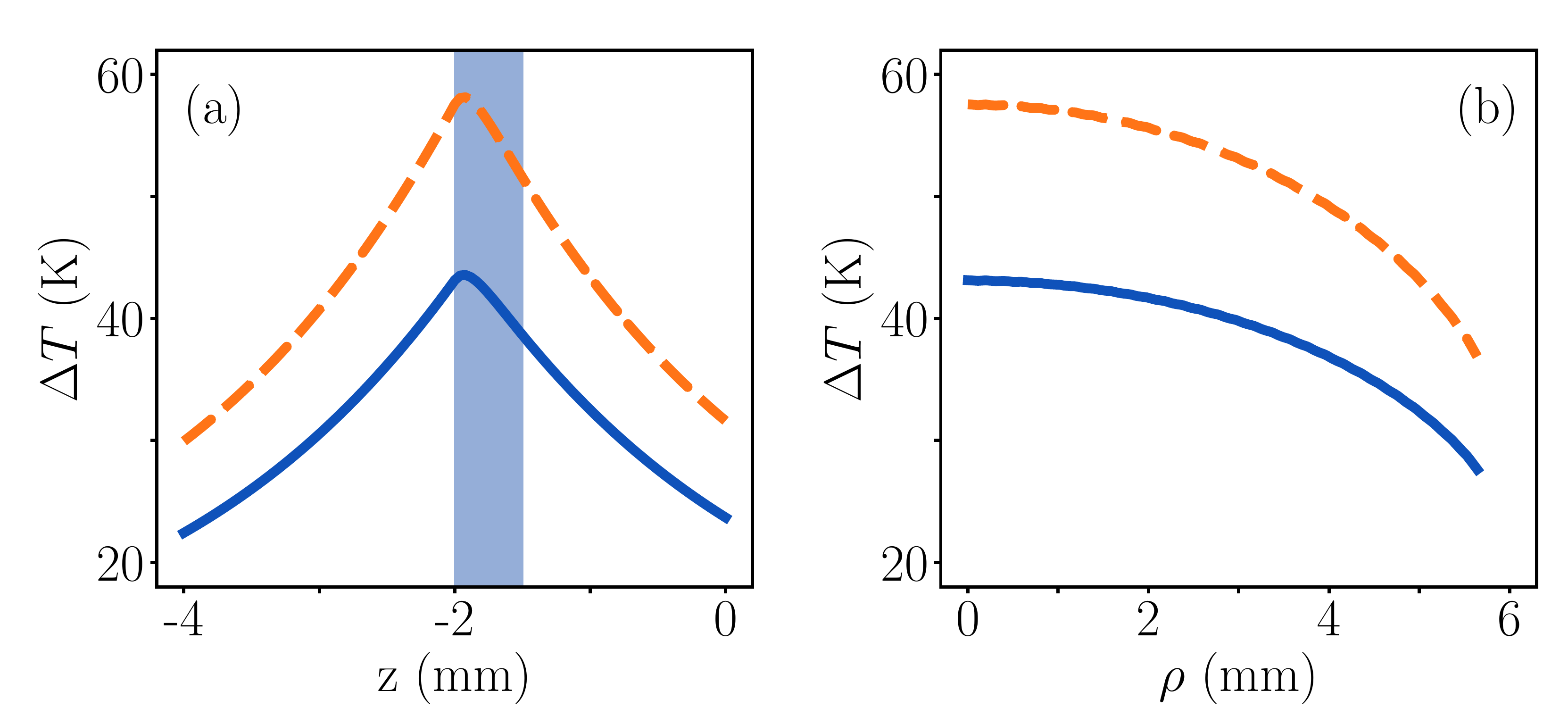}
\caption{(Color online) {\XYZ (a) The temperature rise profile along the illumination direction and (b) the temperature rise profile at the surface facing the light source} {\XYZ when the NP array is immersed in a host with $\kappa_h = 50.3$ mW/(m$\cdot$K) (blue solid line) and when it is immersed in a host with $\kappa_h = 38$ mW/(m$\cdot$K) (orange dashed line)}. The blue-gray region represents the NPs array. All other conditions are the same as %Figure~\ref{fig:dtemp_distrib_rev}(b).%
in Section~\ref{sec:temp_distrib}.
}\label{fig:dtemp_vs_h_au6}%Figure~\ref{fig:cabs_csca_dtemp_au_6nm_H2O_532nm}.}\label{fig:dtemp_vs_h_au6}
\end{figure}

Unlike the weak dependence of the overall temperature rise $\Delta T$ on the parameters discussed in the previous subsections, it exhibits a strong dependence on the thermal conductivity of the host $\kappa_h$; specifically, it is inversely proportional to it (see Eq.~(\ref{eq:mti_NP_sum})). As shown in Figure~\ref{fig:dtemp_vs_h_au6}, {\XYZ the temperature rise profile of the same sample increases by a factor of $\sim 4/3$ when the host is changed from air/oxide ($\kappa_h = 50.3$ mW/(m$\cdot$K)) to CO$_2$/oxide ($\kappa_h = 38$ mW/(m$\cdot$K)), in excellent agreement with Eq.~(\ref{eq:dTtop_CW})}.

\subsection{Pulse train illumination}\label{sec:pulse_illum}
We now turn our attention to the temperature rise dynamics of the catalyst sample under a pulse train illumination and the sensitivity of the results to the various system parameters. We consider a pulse train illumination {\XYZ with (time) average intensity $\langle I_{\textrm{inc}} \rangle = 80$ mW/cm$^2$} (as for the CW case above), pulse repetition rate $f = 80$ MHz, pulse duration $\tau = 4$ ps and peak intensity  $I_0 = \sqrt{\pi/2}\langle I_{\textrm{inc}} \rangle/(\tau f)$. 

During each single pulse event, the inner temperature of each NP increases due to photon absorption, then the inner NP temperature decays due to heat transfer to the host. This heat diffusion from the (many) other NPs keeps the sample warm until all the thermal energy diffuses out of the sample. 

\begin{figure}[h]
\centering
\includegraphics[width=0.5\textwidth]{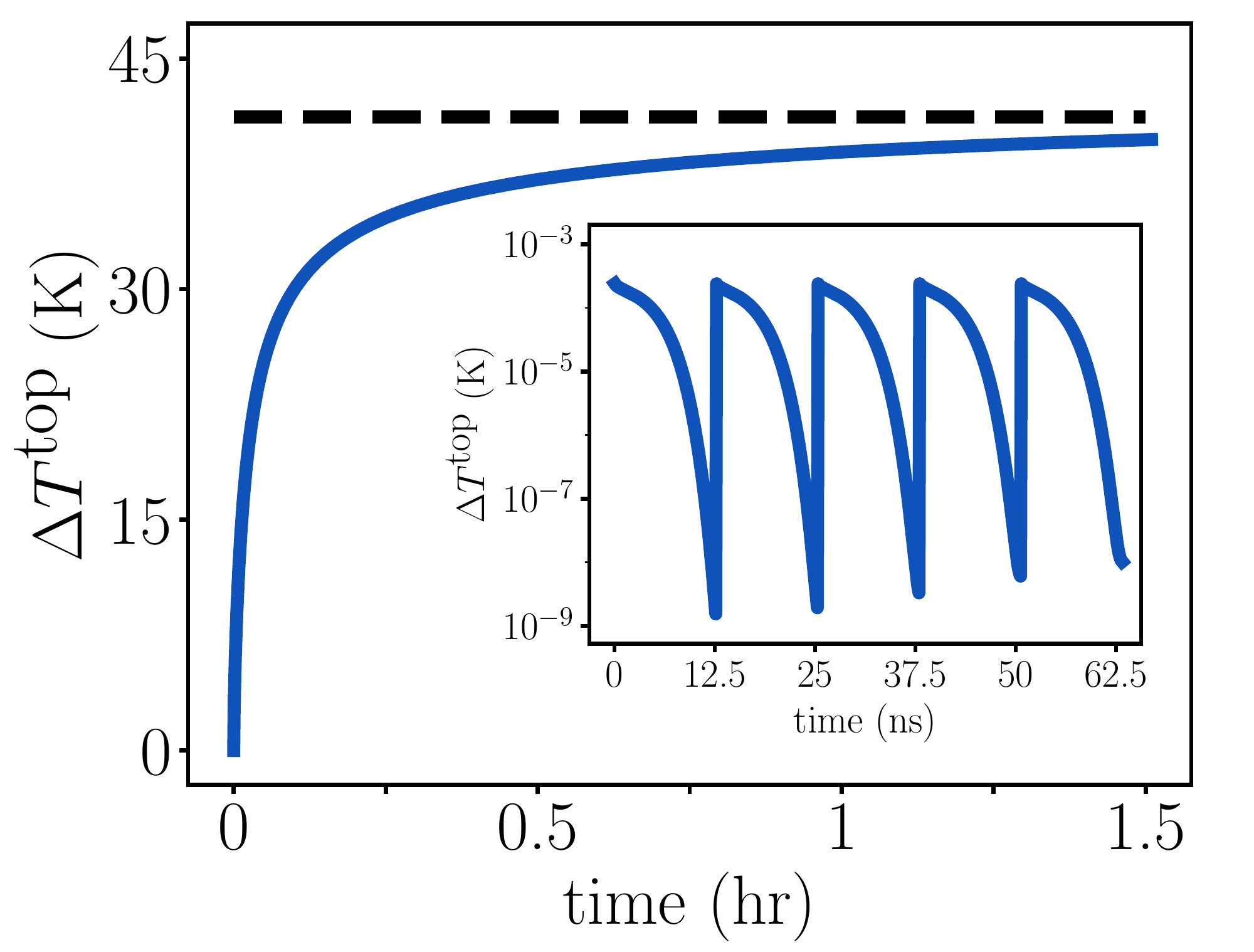}
\caption{(Color online) The temporal evolution of $\Delta T^{\textrm{top}}$ under {\it pulse train} illumination given by Eq.~(\ref{eq:dT_mlt_pls}). The black dashed line represents the ``steady-state'' temperature~(\ref{eq:dTtop_stdy_approx}). All other conditions are the same as %Figure~\ref{fig:dtemp_distrib_rev}(b).
in Sec.~\ref{sec:temp_distrib}. The insert shows the temperature evolution during the illumination of the first several pulses.} \label{fig:dtemp_vs_time_imp_stdy}
\end{figure}

The spatio-temporal evolution of the sample temperature under a pulse train illumination can be obtained by the linear combination of many solutions of (time-shifted) single pulse events (see details in Supplementary Information Section~\ref{app:temp_dyn_multi_pulse}). Since the pulse repetition rate is faster than the overall decay time to the environment, the temperature is increasing in a step-wise fashion. This heat accumulation finally slows down and the temperature reaches a ``steady-state'' at $\sim$ 45 K on a time scale of a few minutes to a few hours, as shown in Figure~\ref{fig:dtemp_vs_time_imp_stdy}. This was indeed observed in e.g. \cite{Halas_Science_2018}.

Indeed, we show in Supplementary Information Section~\ref{app:est_dT1_st_pluse} that the ``steady-state'' temperature rise of $\Delta T^{\textrm{top}}_{\textrm{mp}}$ can be approximated by 
\begin{align}\label{eq:dTtop_stdy_approx}
\Delta T^{\textrm{top}}_{\textrm{mp}}(t\rightarrow\infty) \approx\dfrac{\langle I_{\textrm{inc}}\rangle \rho_0}{2\kappa_h}\left( 1 - e^{-H/\delta_{\textrm{skin}}} \right).
\end{align}
This prediction is found to be in excellent agreement with the numerical results, see Figure~\ref{fig:dtemp_vs_time_imp_stdy}. Importantly, pulsed and CW cases give the same result of temperature rise in the ``steady-state'' (compare Eqs.~(\ref{eq:dTtop_CW}) and (\ref{eq:dTtop_stdy_approx})). This is a manifestation of the fact that once the systems reach a ``steady-state'', macroscopic heating is obtained by a balance of the heat generation and the heat diffusion out of the sample as a whole. This again shows that although plasmonic NPs are thought of as a nanoscale heat source, they eventually cause heating which does not differ so much from macroscopic heat sources. Except for relatively large NPs and/or high intensity pulses, the transient NP temperature rise following each inddividual pulse is small with respect to the base line (average) heating~\cite{Halas_Science_2018,Y2-eppur-si-riscalda}. Other exceptions are obviously the early stages of pulsed illumination and the case of a dilute sample (e.g.~\cite{Boltasseva_LPR_2020}). Nevertheless, the heating is generally weak in those scenarios.

Eq.~(\ref{eq:dTtop_stdy_approx}) not only provides a simple way to compute the ``steady-state'' temperature rise in the pulsed case, but also provides insight to the sensitivity of the ``steady-state'' temperature rise to various system parameters. Specifically, (i) the ``steady-state'' temperature rise {\it is independent} of the pulse duration $\tau$ and the repetition rate $f$; (ii) since the ``steady-state'' temperature rise is the same as for the CW case, all other results are valid here too.

\subsection{Comparison of a simplified sample geometry and a realistic one}\label{sec:remark}
{\XYZ In the previous sections, we studied the parametric dependence of a simplified model using a uniform host. For realistic configurations in which the material of the catalyst sample and of its surrounding are usually different, the thermal conductivity can be highly non-uniform. In these cases, it requires numerical methods, such as finite element method, solving the heat equation to achieve accurate temperature distribution.  However, as shown in Supplemental Information Section~\ref{app:comsol_non_uniform_kappa}, the dependence of the energy absorbed by the sample on the particle size, on the inter-particle spacing and on the illumination wavelength are not affected by the non-uniformity of the thermal conductivity. On the other hand, the non-uniformity of the thermal conductivity can affect the dependence of the temperature rise on the beam radius. This is because when the beam size changes, the heat transfer path can be modified by the non-uniform thermal conductivity\footnote{For example, if the edge of the disc-shaped catalyst sample is in contact with a high thermal conductivity material, such as metal, much more heat is dissipated from the edge than from the center when the beam size becomes compatible to the catalyst sample area.}. However, the change of the dependence of the temperature rise on the beam size is usually small when the sample area is sufficiently larger than the beam area, see Supplemental Information Section~\ref{app:comsol_non_uniform_kappa}.}

\section{Comparison to experimental studies of plasmon-assisted photocatalysis}\label{sec:compare_to_exp}

Light harvesting systems (e.g., photocatalysis pellets, water purification samples etc.) are usually designed to be optically-thick for the purpose of absorbing all illumination energy. Thus, the results described above show that generically, the temperature rise is such systems will have {\em a weak sensitivity to the illumination wavelength, pulse duration, particle size and density}. However, in the context of photocatalysis, when the chemical reaction rate is enhanced by the photo-thermal effect~\cite{Dubi-Sivan-Faraday,anti-Halas-Science-paper,Y2-eppur-si-riscalda}, the reaction rate enhancement can become more sensitive to these parameters via the exponential dependence of the reaction rates on the temperature. The level of the enhanced sensitivity depends on the activation energy. 

\begin{figure}[h]
\centering \includegraphics[width=0.7\textwidth]{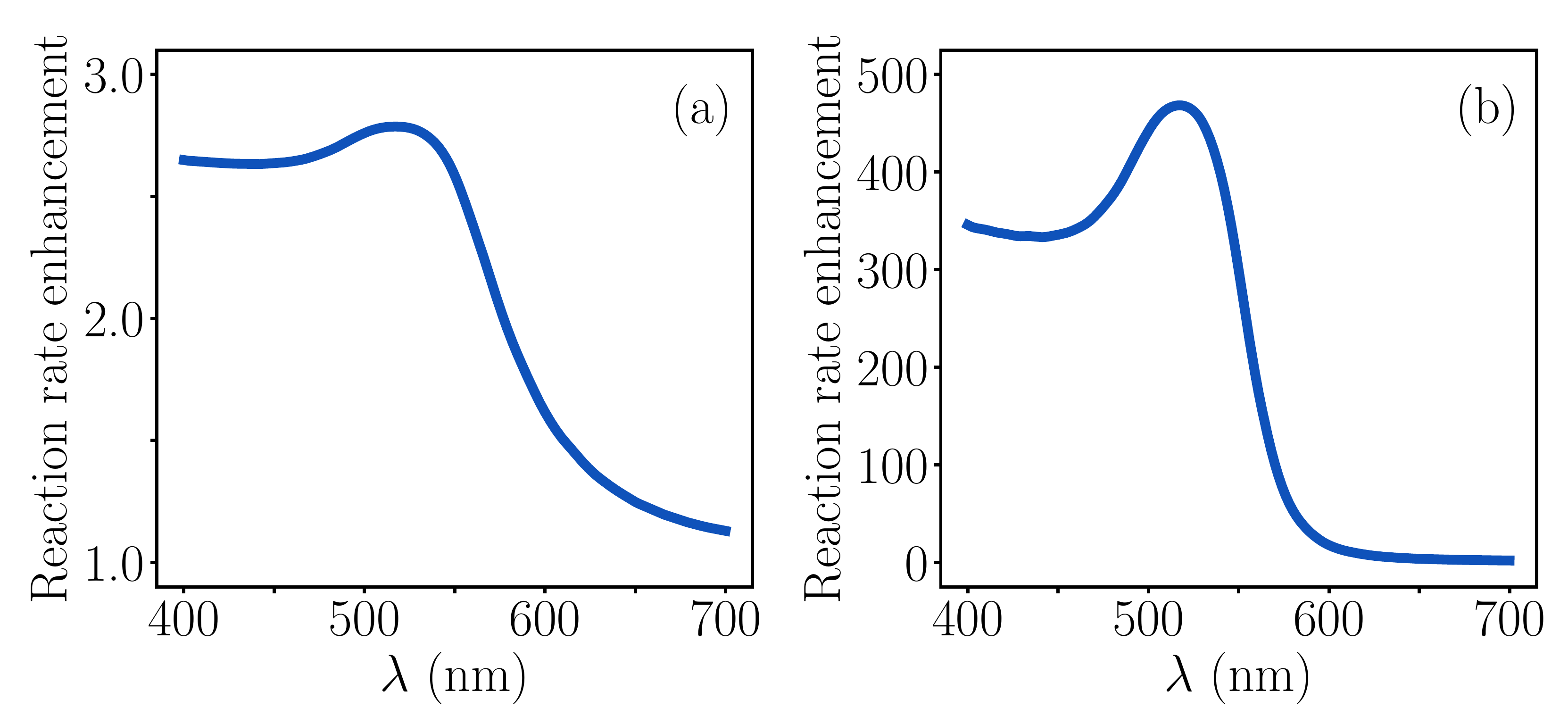}
\caption{(Color online) Reaction rate as a function of the illumination wavelength for activation energy (a) $E_a = 0.2$ eV and (b) $E_a = 1.2$ eV. All conditions are the same as in Figure~\ref{fig:dtemp_vs_wl_au6}.}\label{fig:react_rate_vs_wl_au6_Ea}
\end{figure}

In order to see this, we use as an example the wavelength-dependence of the temperature rise shown in Figure~\ref{fig:dtemp_vs_wl_au6}(b) to investigate the wavelength-dependence of the reaction rate. Specifically, we calculate the reaction rates under illumination for different activation energies ($E_a = 0.2$ eV (as in~\cite{Halas_dissociation_H2_TiO2,Halas_H2_dissociation_SiO2}) and $E_a = 1.2$ eV (as in~\cite{plasmonic_photocatalysis_Linic,Halas_Science_2018})) by using the temperature-shifted Arrhenius Law~\cite{Y2-eppur-si-riscalda}
\begin{align}\label{eq:temp_shifted_Arrhenius}
R(I_{\textrm{inc}}) = R_0\exp\left(-\dfrac{E_a}{k_B \left(T_{h,0} + \Delta T(I_{\textrm{inc}})\right)}\right),
\end{align}
where $k_B$ is the Boltzmann constant, $T_{h,0}$ is the host temperature in the dark, and $R_0$ is a constant that depends on the details of the reactants as well as the details of the measurement. For a fair comparison, we plot the reaction rate enhancement as the ratio of the reaction rate under illumination to the reaction rate in the dark. Figure~\ref{fig:react_rate_vs_wl_au6_Ea} shows that the reaction rate enhancement for $E_a = 1.2$ eV is around a few hundreds, much stronger than that for $E_a = 0.2$ eV (only $2 - 3$).  Moreover, the reaction rate enhancement at the plasmonic resonance wavelength shows a much higher peak for $E_a = 1.2$ eV {\XYZ($35 \%$ higher than the short wavelength shoulder)} than that for $E_a = 0.2$ eV {\XYZ(only $5 \%$).} Both cases are much weaker than the absorption peak shown in Figure~\ref{fig:dtemp_distrib_rev}(a), while the latter is only compatible to the peak of the temperature rise shown in Figure~\ref{fig:dtemp_vs_wl_au6}. The difference in the reaction rate enhancement can be well explained by the Arrhenius equation~(\ref{eq:temp_shifted_Arrhenius}) which states that the higher activation energy, the more sensitive to temperature the reaction rate is.

Despite the somewhat greater sensitivity induced by relatively high activation energies, the conclusion of the above analysis is that the spectral dependence of the temperature distribution and reaction rates is much milder compared to the single NP response. In that sense, the experimental reports reveal a somewhat confusing picture - while some of the more careful studies of plasmon-assisted photocatalysis, see e.g.~\cite{Mirkin_Ag_growth,Liu-Everitt-Nano-Letters-2019,JACS_Xiong} reported a weak spectral dependence of the reaction rate, it was common to associate the faster chemical reactions with the plasmon resonance response. These claims originate, at least partially, from photodetection experiments where the non-thermal electrons had to cross a Schottky barrier (as e.g., in~\cite{Moskovits_hot_es,Moskovits_hot_es_water_splitting,Uriel_Schottky,Uriel_Schottky2,Valentine_hot_e_review}). In earlier stages of the research of these problems, there was no clear distinction in the underlying mechanism ascribed to plasmon-assisted photodetection and photocatalysis experiments. However, while the wavelength dependence is obvious in photodetection studies, in the context of photocatalysis experiments the claims on the dominance of the plasmon resonance were rarely quantified. As we shall show below, they were also sometime inaccurate. In particular, as some works use (solar-like) white light sources, one has to take care how the spectral data is presented and interpreted. 

\begin{figure}[h]
\centering
\includegraphics[width=1\textwidth]{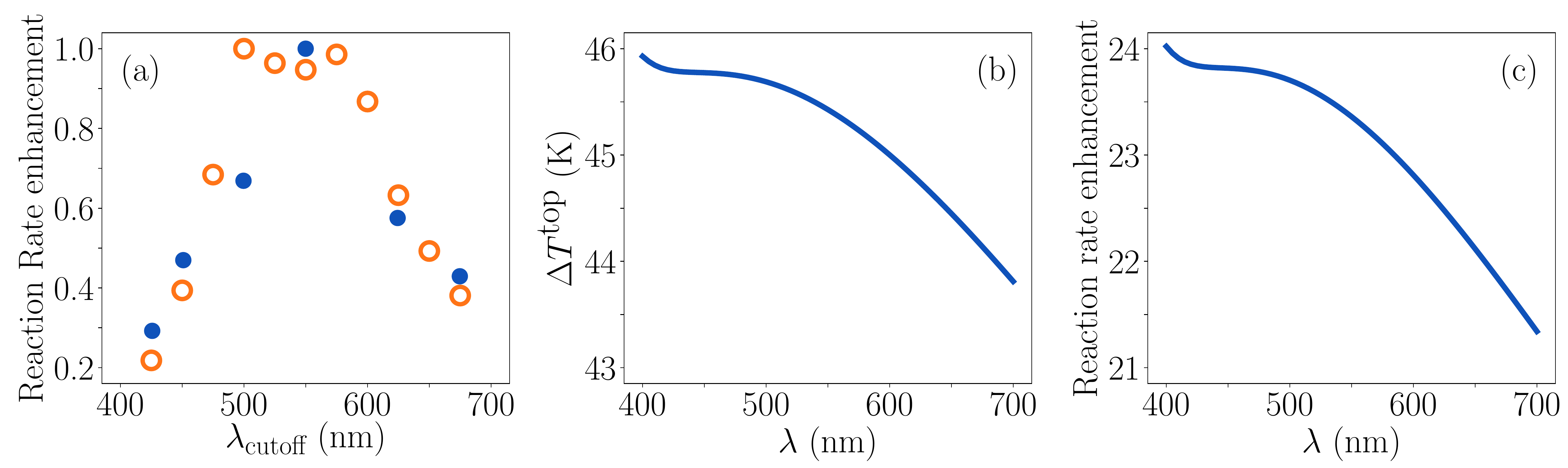}
\caption{(Color online) (a) The spectral dependence of the reaction rate enhancement by a revised thermal calculation following the procedures described in~\cite{Y2-eppur-si-riscalda,plasmonic_photocatalysis_Linic} for the photocatalyst sample used in~\cite{plasmonic_photocatalysis_Linic} (orange circles). The blue dots represent the experimental data copied from~\cite{plasmonic_photocatalysis_Linic}. (b) The wavelength-dependence of the temperature rise at the center of the top surface of the pellet calculated with a (``tunable'') CW light source. For each wavelength, the illumination intensity is 250 mW/cm$^2$ and the bottom of the pellet is fixed to $448$ K, i.e., the measured temperature. (c) The wavelength-dependence of the reaction rate calculated from the temperature rise in (b) by using the temperature-shifted Arrhenius Law~(\ref{eq:temp_shifted_Arrhenius}).}\label{fig:rate_vs_wl}
\end{figure}

To see this, let us consider the work of Christopher~{\it et al}.~\cite{plasmonic_photocatalysis_Linic}, where the wavelength-dependent reaction rate measurements were performed at constant illumination intensity (250 mW/cm$^2$) of a white light source. Instead of using bandwidth limited sources (as e.g., in~\cite{Halas_dissociation_H2_TiO2,Halas_H2_dissociation_SiO2,Liu-Everitt-Nano-Letters-2019}), the authors of~\cite{plasmonic_photocatalysis_Linic} {\XYZ measured the reaction rate for 7 different spectral bandwidths obtained by sending the illuminating light through a series of 7 long pass filters. The cutoff wavelength (the shortest wavelength at which the light can pass through) of these pass filters are, respectively, $\lambda_1 = 400$ nm, $\lambda_2 = 425$ nm, $\lambda_3 = 450$ nm, $\lambda_4 = 500$ nm, $\lambda_5 = 550$ nm, $\lambda_6 = 625$ nm and $\lambda_7 = 675$ nm (Here we denote the corresponding measured reaction rates as $\tilde{R}(\lambda_n)$). Then they plotted the reaction rate difference $(\tilde{R}(\lambda_n) - \tilde{R}(\lambda_{n-1}))/(\lambda_n - \lambda_{n-1})$ as a function of $\lambda_n$.} %measured the reaction rate for 7 different spectral bandwidths obtained by sending the illuminating light through a series of 7 long pass filters. Then, they plotted the reaction rate as a function of the spectral difference 
%{\bf IW - try to explain this in more detail... what was subtracted from what, and how was the x-axis set? you refer to a cutoff wavelength - better define what this is ,,,}. This yielded a prominent spectral peak around $550$ nm, see Figure~\ref{fig:rate_vs_wl}(a), 
{\XYZ which corresponds to that of the light source used in that experiment. Peculiarly, however, the spectral peak of the Ag NPs used in this experiment occurs at much shorter wavelengths ($\sim 430$ nm, see the Supplemental Information in~\cite{Y2-eppur-si-riscalda}). 

In order to explain this observation, we revisit the thermal calculations we performed for this structure in~\cite{Y2-eppur-si-riscalda} and follow the procedure described in~\cite{plasmonic_photocatalysis_Linic} to calculate the reaction rate as a function of the cutoff wavelength. Specifically, first, to account of the non-uniform thermal conductivity in the system, we use COMSOL Multiphysics to perform a series of temperature calculations of the sample in which we mimic the experiment by cutting off the photons of the light source with wavelengths shorter than the threshold of the filter used in the measurement. To do that, we extend our formulation Eq.~(\ref{eq:mti_NP_sum}) in Section.~\ref{sec:model} and Eq.~(\ref{eq:pabs_avg}) in Supplementary Information Section~\ref{app:temp_distrib_CW} from the monochromatic to the polychromatic illumination. The intensity of the polychromatic light source $I_{\textrm{inc}}$ is related to its spectrum $i_{\textrm{inc}}(\omega)$ by $I_{\textrm{inc}} = \int i_{\textrm{inc}}(\omega) d\omega$ and the average absorbed power density by the NP at ${\bf r}_i$ in Eq.~(\ref{eq:pabs_avg}) becomes $\bar{p}_{\textrm{abs},i} = \dfrac{1}{V_{\textrm{NP}}}\int i_{\textrm{inc}}(\omega) e^{- z_i/\delta_{\textrm{skin}}(\omega)} \sigma_{\textrm{abs}}(\omega) d\omega$; more details can be found in~\cite{Y2-eppur-si-riscalda}. Next, we apply the temperature-shifted Arrhenius Law~(\ref{eq:temp_shifted_Arrhenius}) to calculate the reaction rate. Finally, we compute the spectral differences of the reaction rate obtained from the temperature-shifted Arrhenius Law~(\ref{eq:temp_shifted_Arrhenius})~\cite{Y2-eppur-si-riscalda} as in~\cite{plasmonic_photocatalysis_Linic} and in the description above. The experimental result and our calculation are shown in Figure~\ref{fig:rate_vs_wl}(a) to provide remarkable agreement; this provides further support to the re-interpretation of this specific experiment in~\cite{Y2-eppur-si-riscalda} as originating from a pure thermal effect. Most importantly, a similar calculation performed with a (``tunable'') CW source reveals a rather shallow spectral dependence for both the sample temperature and reaction rate, and a maximum near the actual plasmon resonance of the Ag NPs used in that experiment, see Figure~\ref{fig:rate_vs_wl}(b)-(c). This shows that the spectral dependence shown in~\cite{plasmonic_photocatalysis_Linic} is a result of the measurement procedure and apparatus rather than an intrinsic property of the sample, and that the actual spectral response of that system was flat, in correlation with our claims in Section.~\ref{sec:wavelength_dep_T}.} %{\bf IW - complete please...}. }% All the above shows that one has to be careful when attempting to draw conclusions about the physical origin of the reaction rate based on the spectral characteristics of the system.

\section{Outlook}\label{sec:outlook}
The approach adopted in the current study represents a minimal benchmark for the evaluation of thermal effects in light harvesting systems. It also shows that many of previous claims on parametric dependence etc. may have been inaccurate, and need to be re-evaluated.

Our results contribute further insights to the important task of distinguishing between the roles of thermal and non-thermal effects in plasmon-assisted photocatalysis experiments~\cite{Dubi-Sivan-Faraday,Khurgin-Faraday-hot-es,anti-Halas-Science-paper,Y2-eppur-si-riscalda,R2R,Baldi-ACS-Nano-2018,Liu-Everitt-Nano-Letters-2019,Baffou-Quidant-Baldi}. This distinction is of great importance because if thermal effects are dominant in a specific experiment, then, unlike the claims advocated originally, the use of plasmonic NPs to catalyze the reaction would suffer from all the known limitations of thermal effects in the context of photocatalysis.

First, the identified temperature nonuniformity shows that a standard normalization of the reaction rates in photocatalysis by the catalyst mass (as e.g., in Refs.~\cite{Halas_Science_2018,Halas-Nature-Energy-2020}) can incur severe errors in evaluation of the reaction enhancement rate, thus, invalidating the conclusions of these papers, especially since these studies employed different samples for experiment and control; see also discussion in~\cite{R2R}. This is particularly problematic for the latter work, which was published after this specific criticism was brought to the attention of its authors. As shown in~\cite{Liu-Everitt-Nano-Letters-2019}, even a more careful procedure to extract an effective temperature for the sample (via recursive evaluations of the reaction rate and the sample temperature may not be sufficient to explain the intricate details of the reaction rate enhancement.

Second, while both non-thermal (``hot'' carrier) and thermal effects exhibit a similar dependence on the illumination intensity and the absorption cross-section, the latter exhibits a much stronger dependence on the parameters of the system (NP size and shape, density, illumination wavelength, sample thickness etc.), at least for the typical optically-thick samples. Similarly, thermal effects exhibit a sensitivity to the thermal properties of the host which non-thermal effects naturally lack. Thus, variations of the various parameters may hint toward the mechanism underlying the enhanced chemical reactions studied. These ideas can complement the simple experimental tricks suggested already in~\cite{Baffou-Quidant-Baldi} towards the same goal.

Having said the above, we should recall that in the weak illumination limit (to which the majority of plasmon-assisted photocatalysis experiments conform), thermal effects greatly dominate non-thermal effects. This was shown for a single NP in~\cite{Dubi-Sivan}, and is obviously more pronounced for NP ensembles for which the thermal effects accumulate, but non-thermal effects remain the same on the level of each individual NP.

Overall, the calculations performed here are simple, but were not performed so far in the current context, at least not systematically. They can however be extended to more complicated scenarios. For example, our model can be used to study the transient temperature rise of the sample and the temporal evolution of the reaction rate. This is important when the illumination time is shorter than the time scale required for the system to reach the steady-state, especially for catalyst samples immersed in host material with low diffusivity. 

Our model can also be extended to account for heat convection by gas or liquid flow. However, as shown in~\cite{Y2-eppur-si-riscalda,Dubi-Sivan-APL-Perspective}, under realistic conditions, these effects provide only a modest level of homogenization of the temperature. These explicit calculations show that claims for uniform temperature profiles raised in e.g.,~\cite{Halas_Science_2018,Naldoni-Govorov-2020} are likely to be incorrect, especially since they are not based on an actual calculation or estimate. 

Finally, we note that when the temperature rise is greater than 100 K, it is necessary to take into account the temperature dependence of the optical and thermal properties of the metal and the host material (see e.g.,~\cite{japanese_size_reduction,Sivan-Chu-high-T-nl-plasmonics,Gurwich-Sivan-CW-nlty-metal_NP,IWU-Sivan-CW-nlty-metal_NP}). The latter is expected to have a significant effect on the increase of the sample temperature and of the reaction rate, see discussion in~\cite{R2R}. A complete model that include temperature-dependent parameters will have to be left for a future study.

\section*{Conflicts of interest}
There are no conflicts to declare.

\section*{Acknowledgements}
The authors would like to thank Y. Dubi for many useful discussions. YS and IWU were partially supported by Israel Science Foundation (ISF) grant no. 899/16.

\section*{}
\noindent {\bf Electronic Supplementary Information (ESI) available:} Steady state temperature distribution under CW illumination (Section~\ref{app:temp_distrib_CW}), An estimate of the steady-state $\Delta T^{\textrm{top}}$ under CW illumination (Section~\ref{app:est_dT1_st_CW}), Spatio-temporal evolution of the sample temperature under a pulse train illumination (Section~\ref{app:temp_dyn_multi_pulse}), and An estimate of the steady-state temperature rise $\Delta T^{\textrm{top}}$ under pulse train illumination (Section.~\ref{app:est_dT1_st_pluse}). See DOI: 00.0000/00000000.

%\bibliographystyle{rsc}
%\bibliography{my_bib_2}

\providecommand*{\mcitethebibliography}{\thebibliography}
\csname @ifundefined\endcsname{endmcitethebibliography}
{\let\endmcitethebibliography\endthebibliography}{}

\end{document}